\documentclass[twocolumn]{aastex631}
\usepackage{makecell}
\usepackage{amsmath}
\usepackage{xcolor}

\begin{document}

\title{High-redshift JWST massive galaxies and the initial clustering of supermassive primordial black holes}

\author[0009-0003-1218-2569]{Hai-Long Huang}
\affiliation{School of Fundamental Physics and Mathematical
    Sciences, Hangzhou Institute for Advanced Study, UCAS, Hangzhou
    310024, China}
\affiliation{School of Physical Sciences, University of
Chinese Academy of Sciences, Beijing 100049, China}
\email{huanghailong18@mails.ucas.ac.cn}

\author[0000-0003-0957-3633]{Jun-Qian Jiang}
\affiliation{School of Physical Sciences, University of
Chinese Academy of Sciences, Beijing 100049, China}
\email{jiangjq2000@gmail.com}

\author{Yun-Song Piao}
\affiliation{School of Fundamental Physics and Mathematical
    Sciences, Hangzhou Institute for Advanced Study, UCAS, Hangzhou
    310024, China}
\affiliation{School of Physical Sciences, University of
Chinese Academy of Sciences, Beijing 100049, China}
\affiliation{International Center for Theoretical Physics
    Asia-Pacific, Beijing/Hangzhou, China}
\affiliation{Institute of Theoretical Physics, Chinese
    Academy of Sciences, P.O. Box 2735, Beijing 100190, China}
\email{yspiao@ucas.ac.cn}

\begin{abstract}

In this paper, we show that the initial clustering of supermassive
primordial black holes (SMPBHs) beyond a Poisson distribution can
efficiently enhance the matter power spectrum, and thus the halo
mass function. As a result, the population of initially clustered
SMPBHs with $M_{\rm PBH}\sim 10^9M_\odot$ and the fraction of
energy density $f_{\rm PBH}\sim 10^{-3}$ (consistent with current
constraints on SMPBHs) has the potential to naturally explain
high-redshift massive galaxies observed by the James Webb Space
Telescope.

\end{abstract}



\section{Introduction} \label{sec:intro}

Recent observations with the James Webb Space Telescope (JWST)
have discovered a large number of quasars powered by supermassive
black holes (SMBHs) already in place within the first few hundred
million years after the Big Bang
\citep[e.g.][]{2023ApJ...959...39H,2023ApJ...953L..29L,2023arXiv230801230M,
2023arXiv230605448M,2023A&A...677A.145U,2023ApJ...955L..24G,
2023ApJ...957L...7K,2023arXiv231203589U,2023arXiv230805735F,
2023arXiv230512492M,2024NatAs...8..126B,2024ApJ...964...39G,
2024ApJ...960L...1N,2024ApJ...965L..21K}. The existence of such
SMBHs at very high redshifts poses a significant challenge. It is
commonly assumed that SMBH seeds initially form, such as seed
black holes expected from Pop
\uppercase\expandafter{\romannumeral3} stellar remnants
\citep{2001ApJ...551L..27M}, and then grow into SMBHs over time
until the redshifts of observations. However, even if they accrete
at the maximal Eddington rate, there does not seem to be enough
cosmic time to grow them sufficiently.\footnote{It might be
thought the hypothesis that heavy seeds could be direct collapse black holes
(DCBHs) might alleviate this problem. However, DCBHs have long been
considered too rare to account for the entire observed SMBH
population, see also \citet{2024arXiv240614658B} for recent
numerical simulations.} In this scenario, the direct birth of
SMBHs in the very early Universe, i.e., SMPBHs, becomes rather
appealing.

The initial clustering of stellar-mass PBHs has been ruled out by
microlensing and Lyman-$\alpha$ forest observations
\citep{2022PhRvL.129s1302D}. However, it seems that SMPBHs prefer
an initially clustered spatial distribution. In the standard
mechanism that PBHs are from a direct collapse of large amplitude
perturbations after horizon entry \citep[e.g.][]{
2018CQGra..35f3001S,2021RPPh...84k6902C,2021JPhG...48d3001G,
2022arXiv221105767E,2024NuPhB100316494G}, non-Gaussian primordial
perturbations seems to necessary to evade the $\mu$-distortion
constraints for SMPBHs \citep{2018PhRvD..97d3525N,2021PhRvD.103f3519A,
2023EL....14249001G,2024JCAP...04..021H,2024arXiv240418475B,
2024arXiv240418474S,2024PhRvL.133b1001C}, while the
non-Gaussianity of primordial perturbation will possibly lead to
the initial clustering of SMPBHs
\citep{2018JCAP...03..016F,2015PhRvD..91l3534T,2015JCAP...04..034Y,
2019PTEP.2019j3E02S}. In the mechanism
\citep{2023arXiv230617577H,2023arXiv231211982H} where SMPBHs are
sourced by the inflationary bubbles and have a multi-peak mass
spectrum, the PBHs can be spatial-clustered naturally in
multistream inflation
\citep{2009JCAP...07..033L,2009PhRvD..80l3535L}, see also
\citet{2023arXiv231211982H,2019PhRvD.100j3003D,2021PhLB..81636211L}.
The PBHs from Affleck-Dine mechanism also show a strong clustering
\citep{2021JCAP...10..025K,2024arXiv240509790K}, see also
\citet{2023JCAP...10..049K}.

It has been showed in
\citet{2023arXiv230617577H,2023arXiv231211982H,
2023arXiv230701457G,2023arXiv230617836D,2024PhRvD.109f3515H} that
the merger of SMPBHs might be the source of nano-Hertz
gravitational wave background recently detected by pulasr timing
arrays (PTA) collaborations
\citep{2023ApJ...951L...8A,2023RAA....23g5024X,2023ApJ...951L...6R,
2023A&A...678A..50E}, however, the distribution of SMPBHs must
exhibit some clustering to magnify the merger rate of SMPBH
binaries. On the other hand, considering the non-zero angular
auto-correlation function of quasars
\citep{1988MNRAS.230P...5E,1989MNRAS.237.1127C}, models predicting
the Poisson distribution of SMPBHs are ruled out. The Sloan
Digital Sky Survey \citep{2000AJ....120.1579Y} and the 2dF
QSO redshift survey \citep{2004MNRAS.349.1397C} have
measured the auto-correlation function of quasars up to
$z\approx4$ 
\citep[e.g.][]{2024MNRAS.528.4466P,2024arXiv240312140P}. The
auto-correlation function of faint quasars at $z\approx6$ was also
recently estimated, benefiting from the high sensitivity of the
Subaru High-$z$ Exploration of Low-Luminosity Quasars 
survey \citep{2023ApJ...954..210A}.

Recently, based on 14 galaxy candidates with masses in the range
$10^9-10^{11}M_\odot$ at $7<z<11$ identified in the JWST 
Cosmic Evolution Early Release Science
program, \citet{2023Natur.616..266L} found that the cumulative
stellar mass density at redshift $z^{\rm obs}=8$ for the stellar 
mass density $M_*\ge10^{10}M_\odot$ is
\begin{equation} \label{rhobs}
\rho^{\rm
obs}(M_*\ge10^{10}M_\odot)\simeq1.3\times10^6\rm~M_\odot{\rm
Mpc}^{-3}.
\end{equation}
This result is hardly reconcilable with the 
standard $\Lambda$ Cold Dark Matter model ($\Lambda$CDM)
expectation, which would require an implausible high star
formation efficiency, even larger than the cosmic baryon mass
budget in collapsed structures. This conflict has inspired lots of
relevant studies
\citep[e.g.][]{2023ApJ...944..113B,2023PhRvD.107d3502H,2023PhRvD.108d3510J,2023arXiv230309391Y,2023MNRAS.526L..63P,2023ApJ...953L...4P,2023JCAP...10..072A,2023arXiv231115083D,2024JCAP...05..097F,2024arXiv240112659M,2024PhRvD.109j3522P,2024PDU....4401496I}.

\begin{figure}[ht!]
    \includegraphics[width=\columnwidth,clip]{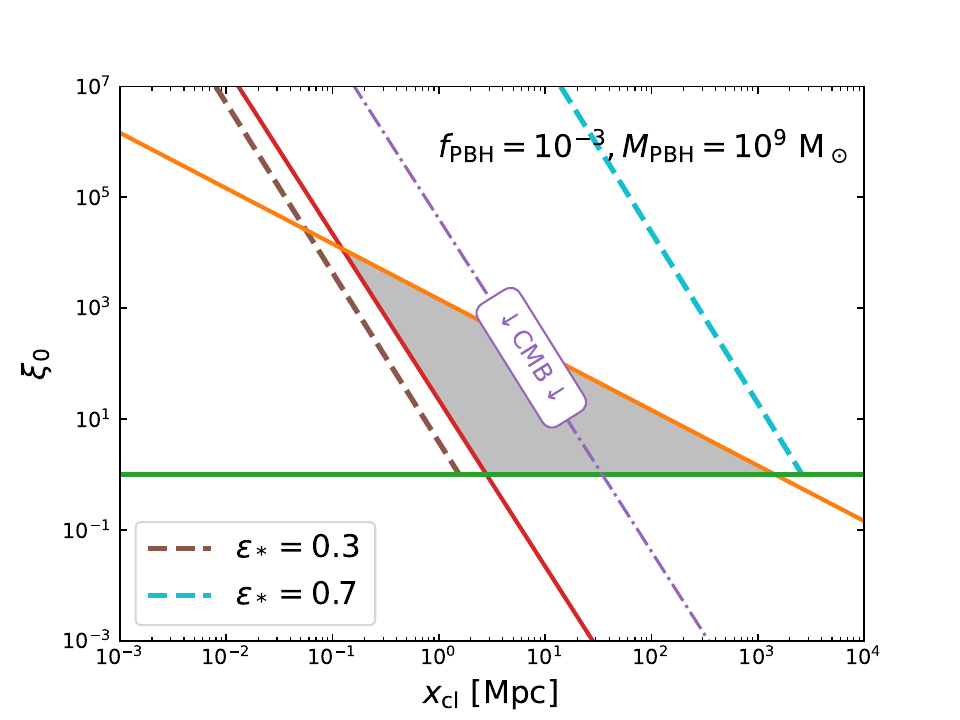}
\caption{The parameter space of $(\xi_0,x_{cl})$ for initially clustered
SMPBHs explaining the JWST observation, where we have
assumed $M_{\rm PBH}=10^9~\rm M_\odot$ and $f_{\rm PBH}=10^{-3}$.
The gray shaded region represents the parameter space allowed by
the model itself, constrained by Eq.~\eqref{eq:xi0dayu1} (green),
Eq.~\eqref{eq:lim1} (red) and Eq.~\eqref{eq:lim2} (orange),
respectively. The purple line represents the observation
constraint from the CMB, Eq.~\eqref{eq:limCMB}. The brown (cyan)
line indicate clustering SMPBH which can explain the comoving
cumulative stellar mass density for the star formation rate
$\epsilon_*=0.3$ (0.7).  
We do not consider the enhancement of the 
PBH isocurvature power 
spectrum at $k>k_{\rm cut,cluster}$ 
(see Eq.~\eqref{eq:kcut} and the discussion nearby), and any
correlation between the adiabatic and 
isocurvature modes is ignored here. Therefore, the estimates of the required
star formation efficiency shown here are the most conservative results
in clustered SMPBH model.}
    \label{fig:const1}
\end{figure}

However, if SMPBHs do exist, they would enhance the halo mass
function and thus boost the early formation of massive galaxies
\citep[e.g.][]{2022ApJ...937L..30L,2023arXiv231204085L,
2024A&A...685L...8C,2024arXiv240511381Z,2023ApJ...944..113B}. In
this paper, we explore the effect of the initial clustering of
SMPBHs. It is found that the initial clustering can further
enhance the halo mass function at the high mass tail. As a result,
the population of initially clustered SMPBHs with $M_{\rm
PBH}\sim 10^9M_\odot$ and $f_{\rm PBH}\sim 10^{-3}$ (consistent
with current constraints on SMPBHs e.g.
\citet{2024PhR..1054....1C}) has the potential to naturally
explain high-redshift massive galaxies observed by JWST. The
results are summarised in Fig.~\ref{fig:const1}.

\section{Modelling initial clustering of PBHs}
\label{sec:model}

The clustering of PBHs beyond the Poisson distribution is
described by its correlation functions, and the precise shape is
model-dependent. Here, for simplicity the correlation function of
PBHs is modelled as \citep{2022PhRvL.129s1302D}
\begin{align}\label{eq:xix}
    \xi_{\rm PBH}(x)=
\begin{cases}
    \xi_0  \qquad {\rm if} \qquad x\le x_{\rm cl},  \\
    0 \qquad  \ \, {\rm otherwise},
\end{cases}
\end{align}
where $x_{\rm cl}$ is the comoving size of the gravitationally
bound cluster. The corresponding PBH clusters follow a Poisson
distribution at much large scales.


It has been argued in \cite{2018PhRvD..98l3533D} that PBHs are
anti-correlated at short distances as one should expect that there
is at most one PBH per horizon volume, i.e. $\xi_{\rm
PBH}(x)\simeq-1 \quad {\rm for} \quad x\lesssim x_{\rm exc}$,
where the comoving size $x_{\rm exc}$ of the small-scale exclusion
volume is approximately the comoving Hubble radius at formation
time of PBHs,
\begin{equation}
    x_{\rm exc}\simeq x_{H}\simeq\left(\frac{M_{\rm PBH}}{M_\odot}\right)^{1/2}\frac{1}
    {4\times10^6~{\rm Mpc}^{-1}}.
\end{equation}
We have verified that the spatial exclusion condition has
negligible effect on our subsequent calculations ($x_{\rm exc}\ll
x_{\rm cl}$), allowing us to safely adopt Eq.~\eqref{eq:xix}.

Here, for analytical viability, we adopt a monochromatic PBH mass
function with a mass parameter $M_{\rm PBH}$ and the fraction of
dark matter (DM) $f_{\rm PBH}=\Omega_{\rm PBH}/ \Omega_{\rm DM}$. The average
energy density of PBHs is $\bar{\rho}_{\rm PBH}=\bar{n}_{\rm PBH}
M_{\rm PBH}$, where the average (comoving) number density is
\begin{equation}
    \bar{n}_{\rm PBH}\simeq3.3\times10^{10}f_{\rm PBH}\left(\frac{M_{\rm PBH}}{\rm M_\odot}\right)^{-1}~{\rm
    Mpc}^{-3}.
\end{equation}
In such a PBH cluster, the number of PBHs is
\begin{equation}
N_{\rm cl}=1+\bar{n}_{\rm PBH}\int{\rm d}^3x\xi_{\rm PBH}(x),
\end{equation}
see \cite{2022PhRvL.129s1302D} for details, which is approximately
$N_{\rm cl} \simeq\frac{4\pi}{3}
    \bar{n}_{\rm PBH}\xi_0x_{\rm cl}^3$. Thus we have that
\begin{equation} \label{eq:xi0dayu1}
    \xi_0\simeq\frac{N_{\rm cl}/\frac{4}{3}\pi x_{\rm cl}^3}{\bar{n}_{\rm PBH}}
\equiv\frac{\bar{n}_{\rm PBH,cl}}{\bar{n}_{\rm PBH}}>1
\end{equation}
represents the ratio of the number density of PBHs within the
cluster to the average number density approximately. The Poisson
distribution corresponds to $\xi_{\rm PBH}(x)=0$, or equivalently,
$N_{\rm cl}=1$.

In clustered scenarios, we should assume that $N_{\rm
cl}\gtrsim3$, i.e.
\begin{equation} \label{eq:lim1}
    \xi_0\gtrsim2.2\times10^{-11}f_{\rm PBH}^{-1}\left(\frac{M_{\rm PBH}}{\rm M_\odot}\right)
    \left(\frac{x_{\rm cl}}{\rm Mpc}\right)^{-3}.
\end{equation}
Additionally, it is required that the cluster does not collapse
into a heavy PBH with mass of $N_{\rm cl}M_{\rm PBH}$, which
suggests \citep{2022PhRvL.129s1302D,2023PhRvL.130q1401D}
\begin{equation} \label{eq:lim2}
    \xi_0\lesssim60f_{\rm PBH}^{-1/3}\left(\frac{C}{200}\right)^{-1/6}\left(
    \frac{x_{\rm cl}}{\rm Mpc}\right)^{-1}
\end{equation}
according to the hoop condition
$r_{\rm cl}\lesssim 2GM_{\rm cl}$, where $C=\mathcal{O}(1-100)$ is a
constant of proportionality, $r_{\rm cl}$ and $M_{\rm cl}$
are the physical radius and mass of the cluster.

\section{Those enhanced by clustering SMPBHs}
\label{sec:structure}


\subsection{Power spectrum}

The density fluctuation of PBHs is
\begin{equation}
    \delta_{\rm PBH}(\textbf{x})\equiv\frac{\delta_{\rm PBH}(\textbf{x})}
    {\bar{\rho}_{\rm PBH}}
    =\frac{1}{\bar{n}_{\rm
    PBH}}\sum_i\delta_D(\textbf{x}-\textbf{x}_i)-1,
\end{equation}
noting that these PBHs are spatially discrete objects. 
$\delta_D(\textbf{x})$ is the three-dimensional Dirac distribution. The
corresponding two-point correlation function of fluctuation is
given by
\begin{equation}
    \langle\delta_{\rm PBH}(0)\delta_{\rm PBH}(\textbf{x})\rangle=\frac{1}
    {\bar{n}_{\rm PBH}}\delta_D(\textbf{x})+\xi_{\rm PBH}(x),
\end{equation}
where $x=|\textbf{x}|$ is the distance between two PBHs.

The power spectrum of the density perturbations of PBHs is
\begin{align}
    P_{\rm PBH}(k)&=\int{\rm d}^3\textbf{x}e^{-i\textbf{k}\cdot\textbf{x}}
    \langle\delta_{\rm PBH}(0)\delta_{\rm PBH}(\textbf{x})\rangle
    \notag \\
    &=P_{\rm Poisson}(k)+P_{\xi}(k),
\end{align}
where $P_{\rm Poisson}(k)=1/\bar{n}_{\rm PBH}$, which is independent of $k$, comes
from the Poisson fluctuation
induced by the fluctuation of the number of PBHs, and
\begin{equation} \label{eq:Pxi}
    P_\xi(k)=\int{\rm d}^3\textbf{x}e^{-i\textbf{k}\cdot\textbf{x}}\xi_{\rm PBH}(x)
    =4\pi\int{\rm d}x x^2\frac{\sin{kx}}{kx}\xi_{\rm PBH}(x)
\end{equation}
comes from the clustered distribution of PBHs. Thus considering
Eq.~\eqref{eq:xix}, we have
\begin{equation} \label{eq:Pxiwithk}
    P_{\xi}(k)=\frac{4\pi\xi_0[\sin{(kx_{\rm cl})}-kx_{\rm cl}\cos{(kx_{\rm cl})}]}{k^3}.
\end{equation}
We observe that
\begin{equation} \label{eq:Pxidomin}
    \lim_{k\to0}P_{\xi}(k)=\frac{4\pi}{3}x_{\rm cl}^3\xi_0\simeq\frac{N_{\rm cl}}{\bar{n}_
    {\rm PBH}}>\frac{1}{\bar{n}_{\rm PBH}}=P_{\rm Poisson},
\end{equation}
which indicates that the contribution of clustering for the power
spectrum $P_\xi(k)$ exceeds that of the Poisson fluctuation on
large scales.

The PBHs contribute to the isocurvature perturbations through $f_{\rm PBH}^2P_{\rm
PBH}(k)$. These isocurvature fluctuations then grow during the matter-dominated era, and
at $z=0$, they are given by
\begin{equation} \label{eq:Piso}
    P_{\rm iso}(k,z=0)=
    \begin{cases}
        \left[f_{\rm PBH}D_{\rm PBH}(0)\right]^2P_{\rm PBH}(k), \quad &{\rm if} \quad k\le k_{\rm cut}, \\
    0, &{\rm otherwise},
    \end{cases}
\end{equation}
where $D_{\rm
PBH}(z)\simeq\left(1+\frac{3\gamma}{2\alpha_{-}}\frac{1+z_{\rm
eq}}{1+z}\right)^{a_{-}}$ with
$\gamma=(\Omega_m-\Omega_b)/\Omega_m$ and
$a_{-}=(\sqrt{1+24\gamma}-1)/4$ is the growth factor of
isocurvature perturbations until today
\citep{2019PhRvD.100h3528I,2023arXiv230701457G}. 
These isocurvature fluctuations are restricted to large scales,
because we expect the linear theory to break down at small
scales. For initially
Poisson-distributed PBHs, the cut-off scale of the spectrum is of
order the inverse mean separation between PBHs \footnote{
Sometimes, a different choice of cut-off scale $k_{\rm
NL}\simeq\left(\bar{n}{\rm PBH}/f_{\rm PBH}\right)^{1/3}$ is
adopted \citep[e.g.][]{2022ApJ...937L..30L,2023arXiv230701457G},
below which the non-linear effects start to dominate. However, for
the SMPBHs with $f_{\rm PBH}\lesssim10^{-3}$, our choice,
Eq.~\eqref{eq:kcut_poisson}, is more conservative. Additionally,
the focus of our investigation is on the effect of the cluster
distribution of PBHs compared to the Poisson distribution, so in
the comparison we only need to maintain the same cut-off
criteria.}
\begin{equation} \label{eq:kcut_poisson}
    k_{\rm cut,Poisson}=(2\pi^2\bar{n}_{\rm PBH})^{1/3}.
\end{equation}
At scales below $k_{\rm cut,Poisson}$,
we will have a single PBH within the corresponding comoving
sphere, signaling the onset of the seed effect prevailing over the
Poisson effect
\citep{2019PhRvD.100h3528I,2023PhRvD.107d3502H,2018MNRAS.478.3756C}.
In the presence of initially clustering, we should replace it with
the inverse average separation of clusters \footnote{We observe
that
\begin{equation}
    \bar{n}_{\rm cl}={\bar{n}_{\rm PBH}\over N_{\rm cl}}<
    \bar{n}_{\rm PBH}\xi_0\simeq\bar{n}_{\rm PBH,cl},
\end{equation}
ensuring that $\bar{n}_{\rm cl}^{1/3}<\bar{n}_{\rm PBH,cl}^{1/3}$,
implying that the average separation between clusters (the minimal
distance between PBHs located in different clusters) is bigger
than that between individual PBHs within a cluster.}
\begin{equation} \label{eq:kcut0}
    k_{\rm cut,cluster}=(2\pi^2\bar{n}_{\rm cl})^{1/3},
\end{equation}
where $\bar{n}_{\rm cl}=\bar{n}_{\rm PBH}/N_{\rm cl}$ is the
cluster number density. It is obvious that it corresponds to
Eq.~\eqref{eq:kcut_poisson} when $N_{\rm cl}=1$ is
satisfied, indicating the self-consistency of our model. Rewriting
it with the model parameters $(x_{\rm cl},\xi_0)$, we have
\begin{equation} \label{eq:kcut}
    k_{\rm cut,cluster}\simeq\left(\frac{3\pi}
    {2\xi_0}\right)^{1/3}\frac{1}{x_{\rm cl}}.
\end{equation}

\begin{figure*}
\centering
    {\includegraphics[width=8.5cm]{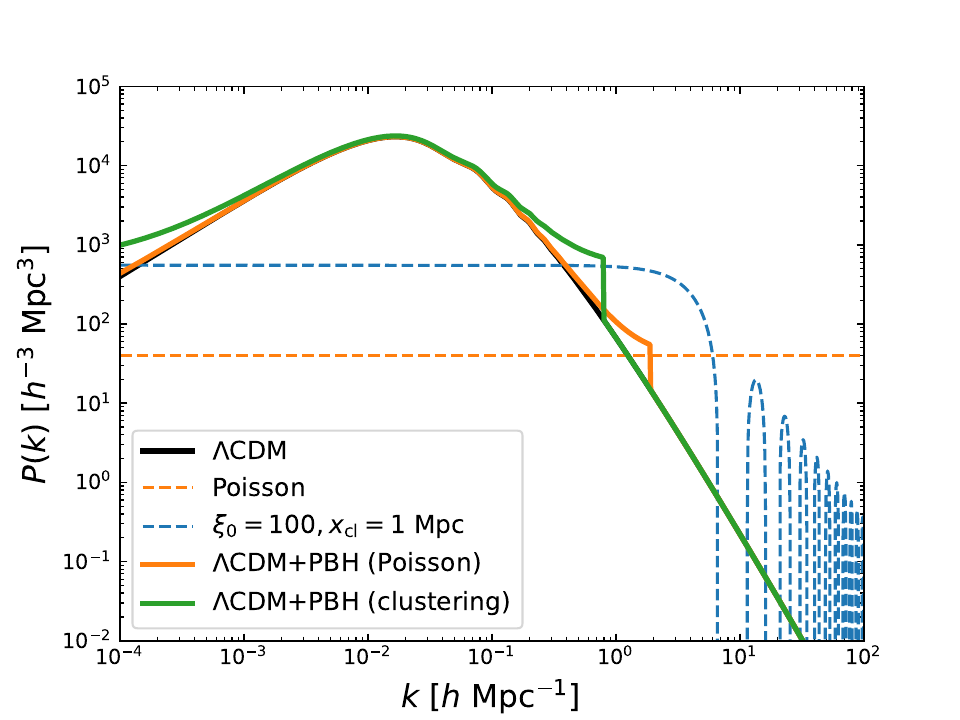}}
    \quad
    {\includegraphics[width=8.5cm]{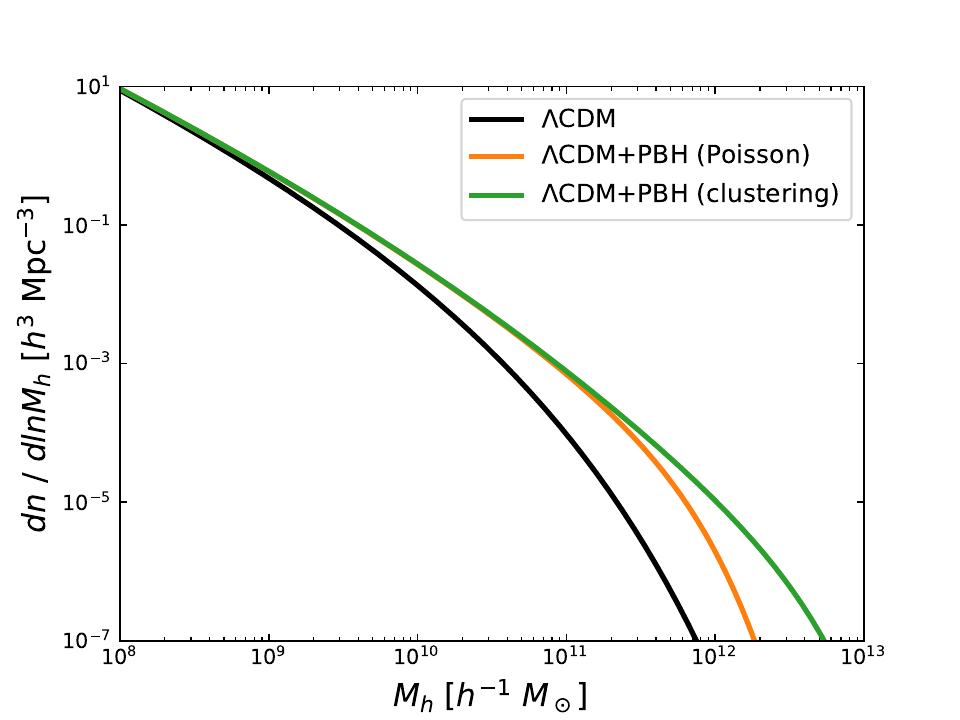}}
\caption{{\it Left panel:} The linear matter power spectrum at
$z=0$ is plotted for $\Lambda$CDM (black solid), for $\Lambda$CDM
with initially Poisson-distributed SMPBHs ($\xi_{\rm PBH}(x)=0$,
orange solid) and initially clustering SMPBHs (green solid),
respectively, where $M_{\rm PBH}=10^9~\rm M_\odot$ and
$f_{\rm PBH}=10^{-3}$. The cut-off Eq.~\eqref{eq:kcut_poisson}
and Eq.~\eqref{eq:kcut} are considered respectively. We also plot
the isocurvature fluctuation sourced by the Poisson fluctuation
(orange dashed) and clustered distribution (blue dashed). {\it Right
panel:} The corresponding halo mass function at redshift $z=8$.}
    \label{fig:power_spectrum_hmf}
\end{figure*}

The Planck results \citep{2020A&A...641A..10P} set the upper bound
on the primordial isocurvature perturbations on the CMB scale as
\begin{equation}
    \beta_{\rm iso}(k_{\rm CMB})\equiv\frac{P_{\rm iso}(k_{\rm CMB})}
    {P_{\rm iso}(k_{\rm CMB})+P_{\mathcal{R}}(k_{\rm CMB})},
\end{equation}
where $P_{\mathcal{R}}(k)=(2\pi^2/k^3)\mathcal{P}_{\mathcal{R}}$
with the power spectrum of curvature perturbations
$\mathcal{P}_{\mathcal{R}}$. We conservatively use the constraint
in \citet{2021JCAP...10..025K}, i.e. $\beta_{\rm iso}<0.035
\quad {\rm at} \quad k_{\rm CMB}=2\times10^{-3}~{\rm Mpc}^{-1}$.
On such a large scale, the isocurvature perturbation
Eq.~\eqref{eq:Piso} can be approximated as
\begin{equation}
    P_{\rm iso}(k)\simeq f_{\rm PBH}^2\times\frac{4\pi}{3}x_{\rm cl}^3\xi_0,
\end{equation}
which suggests
\begin{equation} \label{eq:limCMB}
    \xi_0<4.1\times10^{-2}f_{\rm PBH}^{-2}\left(\frac{x_{\rm cl}}{\rm Mpc}\right)^{-3}.
\end{equation}

\subsection{Halo mass function according to the power spectrum}

We utilize the Sheth-Tormen
\citep{1999MNRAS.308..119S,2001MNRAS.323....1S} modification to
the Press-Schechter formalism \citep{1974ApJ...187..425P} with a
top-hat window function to calculate the halo mass function.
The variance of the smoothed density is
\begin{align}
    \sigma^2(M)&={}\int P_{\rm
    tot}(k,z=0)W^2(kR)\frac{k^2{\rm d}k}{2\pi^2} \nonumber\\
&={}\int \left[P_{\Lambda\text{CDM}}(k,z=0)+P_{\rm
iso}(k,z=0)\right]W^2(kR)\frac{k^2{\rm d}k}{2\pi^2},
\end{align}
where
$W(kR)=\frac{3[\sin{(kR)}-kR\cos{(kR)}]}{(kR)^3}$ is the window
function, 
$R=\left(\frac{3M}{4\pi \rho_m}\right)^{1/3}$
is the comoving radius associated with the mass window $M$, and
$P_{\Lambda\text{CDM}}(k)$ is the power spectrum of adiabatic
perturbations for the $\Lambda$CDM model without SMPBHs.

The peak
significance is defined as
\begin{equation}
    \nu_c(M)=\frac{\delta_c}{\sigma(M,z)}=\frac{\delta_c}{D(z)\sigma(M)},
\end{equation}
where
$D(z)\propto H(z)\int_z^{+\infty}{\rm d}z'\frac{1+z'}{[H(z')]^3}$
is the linear growth function, normalized so that $D(0)=1$. Here,
we adopt $D(z)\approx1.27/(1+z)$, which provides a good
approximation at $z\gtrsim1$ \citep{2024arXiv240313068I}.
The comoving mean number density of halos $n_h(M)$ per logarithmic
mass interval is
\begin{equation}
    \frac{{\rm d}n_h}{{\rm d}\ln M_h}=\frac{\rho_m}{M_h}\nu_c(M_h)f(\nu_c(M_h))
    \frac{{\rm d}\ln \nu_c(M_h)}{{\rm d}\ln M_h},
\end{equation}
where
\begin{equation}
    \nu_cf(\nu_c)=\sqrt{\frac{2}{\pi}}A(p)\left[1+\frac{1}{(q\nu_c^2)^p}\right]
    \sqrt{q}\nu_ce^{-q\nu_c^2/2}
\end{equation}
with $A(p)=\left[1+\pi^{-1/2}2^{-p}\Gamma(0.5-p)\right]^{-1}$.
Here, we adopt $(q,p)=(0.85,0.3)$, which has been found to provide
an accurate fit to the N-body simulations at high redshifts
\citep{2021PhRvD.103h3025S}.

In Fig.~\ref{fig:power_spectrum_hmf}, we depict the power spectrum
and the corresponding halo mass function. The effect of SMPBHs is
creating a small bulge in the matter power spectrum, with the
position and height of the bulge depending on their number
density, while for initial clustering SMPBHs, the bulge shifts to
the left and its height slightly rises. As a
result, the halo mass function is enhanced for
$M_h>10^{11}h^{-1}M_\odot$.

\section{Explain JWST observations}
\label{sec:jwst}


In the following, we explore whether initially clustering SMPBH
can explain the massive galaxy candidates by comparing the number
density of massive galaxies predicted by our SMPBH model with
recent JWST result Eq.~\eqref{rhobs}.

The expected number density of galaxies with $M_*>M_*^{\rm obs}$,
where $M_*^{\rm obs}$ represents the observational threshold of
stellar mass, is given by \citep{2023NatAs...7..731B}
\begin{equation} \label{eq:ngal}
    n_{\rm gal}(M_*\ge M_*^{\rm obs})=\int_{M_h^{\rm cut}}^{\infty}\frac
    {{\rm d}n(z^{\rm obs}, M_h)}{{\rm d}M_h}{\rm d}M_h,
\end{equation}
where $M_h^{\rm cut}=M_h(M_*^{\rm obs})$. Here, we speculate that
each DM halo contains a single central galaxy, thus the relation
between the halo and total stellar mass is
$M_h(M_*)=M_*/(f_b\epsilon_*)$, where $f_b=\Omega_{\rm
b}/(\Omega_{\rm DM}+\Omega_{\rm b})=0.157$ is the baryon fraction
and
$\epsilon_*$ is the star formation efficiency. The comoving
cumulative stellar mass density is
\begin{equation}
    \rho(M_*\ge M_*^{\rm obs})=f_b\epsilon_*\int_{M_h^{\rm cut}}^{\infty}M_h\frac
    {{\rm d}n(z^{\rm obs}, M_h)}{{\rm d}M_h}{\rm d}M_h.
\end{equation}

\begin{figure*}
\centering
    {\includegraphics[width=8.5cm]{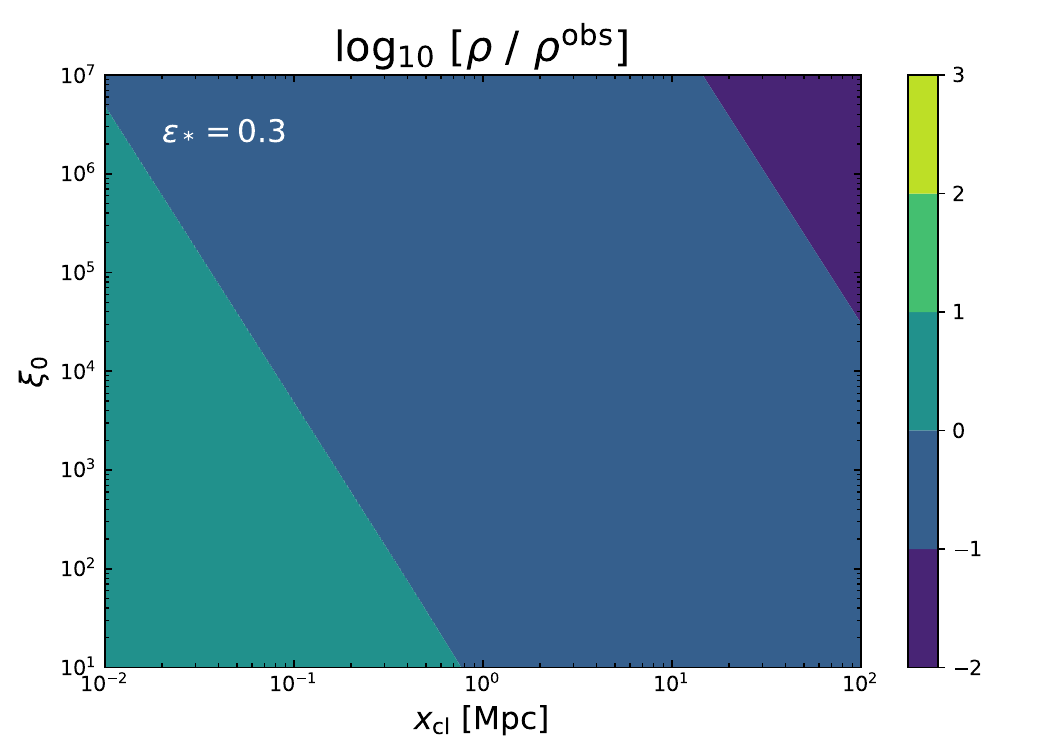}}
    \quad
    {\includegraphics[width=8.5cm]{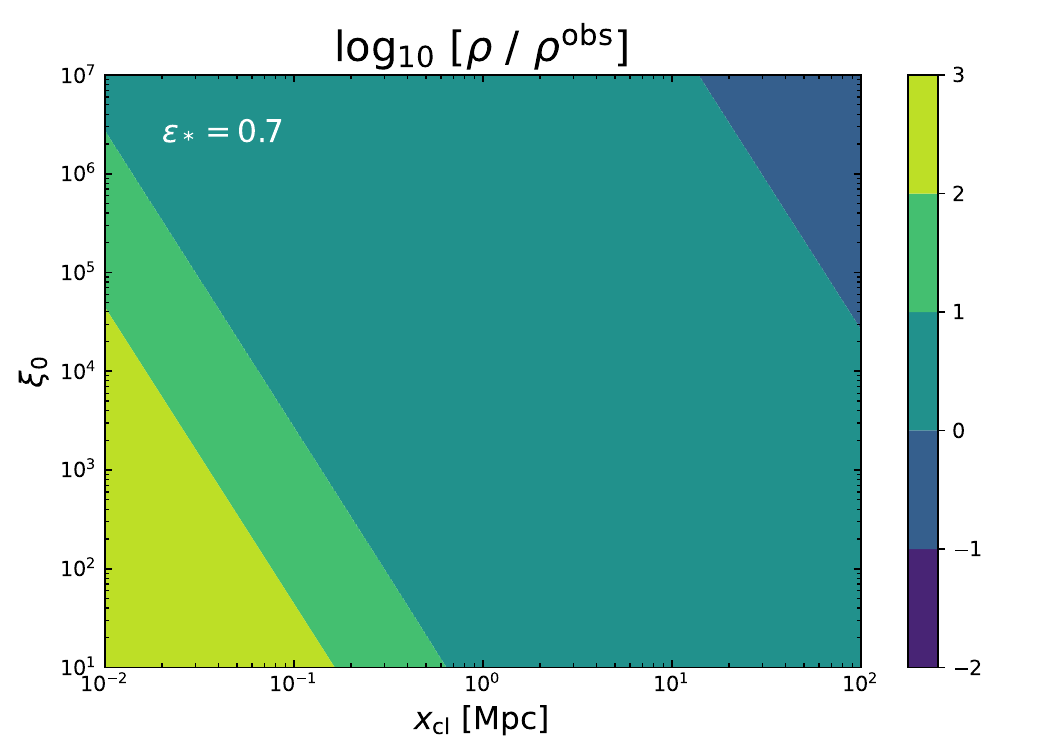}}
\caption{{\it Left panel:} The comoving cumulative stellar mass
density at redshift $z^{\rm obs}=8$ as a function of
$(\xi_0,x_{\rm cl})$, assuming a star formation efficiency
$\epsilon_*=0.3$. {\it Right panel:} Same as the left panel with a
different star formation efficiency $\epsilon_*=0.7$.}
    \label{fig:rho}
\end{figure*}

The comoving cumulative stellar mass density $\rho/\rho^{\rm obs}$
with respect to $(x_{\rm cl}, \xi_0)$ is presented in
Fig.~\ref{fig:rho}, where $\rho^{\rm obs}$ is recent JWST
observation Eq.~\eqref{rhobs}. It can be seen that a population of
clustering SMPBHs has a larger potential to reproduce the
observation results. In contrast, we have
$\log_{10}(\rho/\rho^{\rm obs})\simeq-1.41$ for $\Lambda$CDM model
only and $\log_{10}(\rho/\rho^{\rm obs})\simeq-0.65$ for the same
SMPBHs but with a Poisson distribution, i.e. the $\Lambda$CDM+PBH
(Poisson) model, in the case of $\epsilon_*=0.3$, both are not
sufficient to explain JWST observation, while for a higher
$\epsilon_*=0.7$, we also only have $\log_{10}(\rho/\rho^{\rm
obs})\simeq-0.34$ and $\log_{10}(\rho/\rho^{\rm obs})\simeq0.24$,
respectively. Here, for the $\Lambda$CDM+PBH (clustering) model,
we can naturally have $\log_{10}(\rho/\rho^{\rm obs})\sim 0$ for
$\xi_0=10$, $x_{\rm cl}\sim 1$Mpc and $\epsilon_*\sim 0.3$.

In particular, there exists a parameter space that renders the
model compatible with all constraints on clustering SMPBHs, see
the gray shaded region and the purple line in Fig.~\ref{fig:const1}.
Taking into account these constraints, we observe that the
enhancement of the comoving cumulative stellar mass density caused
by the $\Lambda$CDM with clustering PBHs is about $2$ times larger than
that predicted by Poisson-distributed PBHs. This in certain sense
reflects the advantage of initial clustering SMPBHs in explaining
the high-redshift JWST massive galaxies.

\section{Discussion}
\label{sec:discussion}

High-redshift massive galaxies observed recently by JWST can be
hardly reconcilable with the standard $\Lambda$CDM expectations.
The existence of SMPBHs can enhance the matter power spectrum and
thus the halo mass function, which helps to alleviate this
conflict.

However, SMPBHs can cluster, see \citet{2023arXiv231211982H},
and the merger of SMPBHs initially clustered might be the source
of nano-Hertz gravitational wave background recently detected by
PTA collaboration. This clustering not only sheds light on the
origins of SMBHs but also accelerates the evolution of structures
in the early universe. In this paper, we calculated the effect of
the initial clustering SMPBHs on the matter power spectrum and the
halo mass function, and found that compared with the
Poisson-distributed SMPBHs model, the population of clustering
SMPBHs with $M_{\rm PBH}\sim 10^9M_\odot$ and $f_{\rm PBH}\sim
10^{-3}$ can significantly accelerate the early formation of
massive galaxies, and thus naturally explain recent JWST
observation with a lower star formation efficiency. The collective
results of our analysis is showed in Fig.~\ref{fig:const1}.

The exploration of SMPBHs as potential progenitors of SMBHs that
power high-redshift quasars presents intriguing insights into the
evolution of our Universe. In this case the concept of initially
clustered SMPBHs might be closely related to not only the
evolution of our primordial Universe but also the spatial
distribution of massive objects and the birth of cosmic
structures. The relevant issues are worth of further
investigation.

%

\begin{acknowledgments}
We thank Hao-Yang Liu, Da-Shuang Ye for helpfule discussion. This
work is supported by National Key Research and Development Program
of China (Grant No. 2021YFC2203004), NSFC (Grant No.12075246), and
the Fundamental Research Funds for the Central Universities.
\end{acknowledgments}








\bibliography{refs}{}
\bibliographystyle{aasjournal}



\end{document}